\documentclass[aps,prl,floatfix,twocolumn,footinbib,amsmath,amssymb,reprint]{revtex4-1}

\usepackage{graphics}
\usepackage{epsfig}
\usepackage{natbib}
\usepackage{xspace}
\usepackage{amssymb,amsmath}

\newcommand{\ket}[1]{| #1 \rangle}
\newcommand{\bra}[1]{\langle #1 |}

\renewcommand{\paragraph}[1]{}

\newcommand{\gstate}[6]{\displaystyle
 \left\| _{#1}^{#2}
    \hspace{0cm}_{#3}^{#4}
    \hspace{0cm}_{#5}^{#6}\right\rangle
}

\newcommand{\state}[4]{\displaystyle
 \left\| 
               _{#1}^{#2}
   \hspace{0cm}_{#3}^{#4}
 \right\rangle
}

\newcommand{\sstate}[2]{\displaystyle
 \left\|_{#2}^{#1} \right\rangle
}

\begin{document}

\title{Quantum computation with ultracold atoms in a driven optical lattice}

\author{Philipp-Immanuel Schneider and Alejandro Saenz} 

\affiliation{
         AG Moderne Optik, Institut f\"ur Physik,
         Humboldt-Universit\"at zu Berlin, Newtonstrasse 15,
         12489 Berlin, Germany
}

\date{\today}

\begin{abstract}
We propose a scheme for quantum computation in optical lattices. 
The qubits are encoded in the spacial wavefunction of the atoms such that 
spin decoherence does not influence the computation. Quantum operations are steered by shaking the
lattice while qubit addressability can be provided with
experimentally available techniques of changing the lattice with single-site resolution.
Numerical calculations show possible fidelities above 99\% with gate times on the order of
milliseconds.
\end{abstract}

\maketitle

 In the last years tremendous progress has been made in controlling and observing
 ultracold atoms in optical lattice (OL) potentials~\cite{cold:bloc08a}.
 One of the latest developments has been the optical detection of atoms with 
 single-site resolution in lattices of increasingly smaller periodicity 
 \cite{cold:schr04,cold:kars09,cold:bakr09,cold:sher10}.
 This technological advancement allowed, e.g., for a direct observation of the 
 superfluid to Mott-insulator transition \cite{cold:bakr10}. 
 Along with these detection schemes comes the possibility to control the 
 lattice potential with single-site resolution. Lately, this has been used to
 manipulate the spin of single atoms in an OL \cite{cold:weit11}.

 One of the possibly most important applications of these techniques is the 
 implementation of a quantum computer that would dramatically improve the 
 computational power for particular tasks.
 Compared to other possible candidate systems for the implementation of a quantum 
 computer, neutral atoms in OLs have the advantage of a natural scalability to a 
 large number of atoms encoding qubits and a weak coupling to the environment 
 leading to long decoherence times~\cite{cold:ladd10}.
 Single-site addressability of large qubit systems may be one of the last milestones 
 on the way to an OL quantum computer.

 Many schemes proposed for quantum computation in OLs rely on encoding the qubits by
 atomic spin states~\cite{cold:bren99,cold:haye07,cold:dale08,cold:negr11}.
 Although single-qubit operations \cite{cold:schr04} and collective two-qubit operations 
 \cite{cold:ande07} have been demonstrated, spin qubit states are generally disturbed by 
 external magnetic fields that lead to their decoherence.
 This source of decoherence can be avoided by encoding the qubits in the
 spacial wave function of the ground and first excited state of Bosonic atoms localized 
 at single sites of an OL \cite{cold:ecke02,cold:stra08,cold:negr11}. 
 In this Letter we show that a periodic modulation of the lattice position, i.e. 
 a {\it shaking} of the lattice suffices to perform all quantum operations needed
 for quantum computation. 
 The shaking of the OL can be studied in the context of Floquet theory showing that it 
 effectively changes the hopping parameter of the system~\cite{cold:ecka05}. 
 This effect has been verified experimentally, while revealing that the shaking also 
 drives transitions from the first to the second Bloch band~\cite{cold:lign07}. 
 Qubit addressability is provided by manipulating 
 the OL with single-site resolution. This enables the selective change of the 
 energy spacing of specific sites and the driving of local transitions.
 Additional $z$ rotations induced by the lattice manipulation can be cancelled by
 refocusing schemes known from NMR quantum control~\cite{cold:vand05}.

\begin{figure}[htp]
 \centering
 \includegraphics[width=0.8\linewidth]{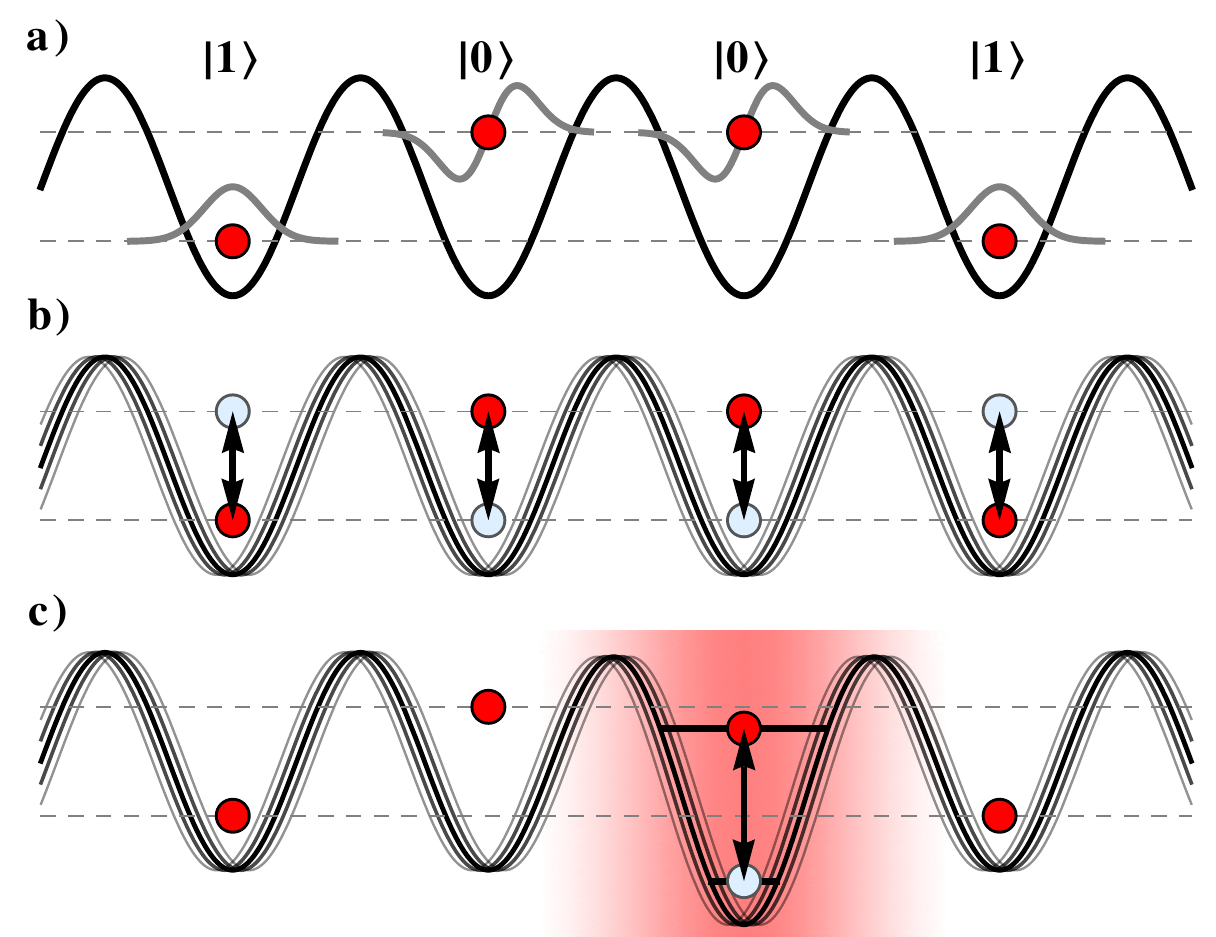}
 \caption{(Color online) 
 {\bf a)} Qubit register with atoms (circles) in the ground ($\sstate{0}{1}=\ket{1}$) or excited state ($\sstate{1}{0}=\ket{0}$) of 
    each lattice site.
 {\bf b)} Shaking of the lattice couples predominantly ground and 
    excited states in each lattice site (indicated by arrows).
 {\bf c)} By shining a laser with waist on the order of the lattice spacing (red shading) 
    onto one site a local transition between $\ket{0}$ and $\ket{1}$ can be driven by 
    shaking the lattice.
}
 \label{fig:system}
\end{figure}

 In the following we consider a one-dimensional (1d) OL in $x$ direction 
 with a potential $V_{L}= V_0 \sin^2(k x)$ assuming tight confinement 
 in the transversal directions. 
 Lengths are given in units of $1/k$ and 
 energies in units of the recoil energy $E_r=\hbar^2 k^2/(2m)$.
 In this system of units the lattice spacing is given as $d=\pi$ and the oscillator
 energy of the harmonic approximation of a single lattice site as $\hbar\omega_{L}=2\sqrt{V_0}$.
 Ultracold Bosons in OLs are regularly described by the Bose-Hubbard model which
 uses the Wannier basis of the first Bloch band to formulate the 
 Hamiltonian in second quantization ~\cite{cold:schn09}. In order to consider excitations 
 to higher Bloch bands the basis has to be extended to include Wannier functions
 $w_{i, b}(x)$ of sites $i=1\dots N$ and several bands $b=1\dots n$.
 Generally, atoms in different Bloch bands are coupled by the interaction
 $V_{\rm int}(x_1,x_2) = g \delta(x_1-x_2)$.
 By means of a Feshbach resonance or strong transversal confinement one can make $g$ 
 sufficiently large to form a Mott insulator state in the first two Bloch 
 bands with one atom per site. A lattice with unit filling then realizes a quantum register with 
 $\ket{0}$ encoded by an atom in the excited state, and $\ket{1}$ encoded by one in the 
 ground state [see Fig.~\ref{fig:system} a)].
 In the following 
\begin{equation}
\label{eq:state_def}
 \left\|
  \begin{array}{cccc}
   n_{2,1} & n_{2,2} & n_{2,3} & \cdots \\
   n_{1,1} & n_{1,2} & n_{1,3} & \cdots \\
  \end{array}
 \right\rangle
\end{equation} 
 shall denote a Fock state in the extended Wannier basis with occupation number 
 $n_{i,j}$ of band $i$ and site $j$.
 If the coupling between the bands and sites is small and no avoided crossing appears all important 
 eigenstates can be characterized by their dominantly contributing Fock state.
 In this notation the qubits state $\ket{001}$ is encoded by the eigenstate 
 with maximal overlap to $\gstate{0}{1}{0}{1}{1}{0}$.

 In the co-moving frame of the lattice a shaking is described by 
 $V_{\rm sh} = f_{\rm sh} x \cos(\omega_{\rm sh} t - \varphi)$,
 where $\omega_{\rm sh}$ is the frequency, $\varphi$ the phase, and $f_{\rm sh}(t)$
 the force amplitude of shaking \footnote{Throughout the Letter we use a pulse 
 envelope $f_{\rm sh}(t)$ that for $0\leq t\leq t_{\rm end}$ is defined by $H_2(\tau)e^{-\tau^2}$ with
 $\tau = (1-2t/t_{\rm end})/\sqrt2$ and $H_2$ the $2^{\rm nd}$ Hermite polynomial.}.
 This perturbation 
 leads to a coupling between 
 Bloch bands with symmetric and antisymmetric Wannier functions [see Fig.~\ref{fig:system} b)]. 
 We note in passing that transitions between the ground and excited states can also be 
 achieved by 
 Raman transitions~\cite{cold:muel07}.
 By means of $\hat V_{\rm sh}$ 
 one is generally able to drive transitions between different eigenstates of the lattice
 Hamiltonian $\hat H_0$.
 For this consider a Hamiltonian $\hat{\mathcal{H}}(t)= \hat H_0 + 
 \lambda \hat V \cos(\omega t - \varphi)$, where $\hat H_0$ has two eigenstates 
 $\ket{\psi_1}$, $\ket{\psi_2}$ with eigenenergies $\hbar\omega_1$ and $\hbar\omega_2$,
 respectively. 
 For $\omega = \omega_2 -\omega_1$ the full time evolution operator in the rotating
 frame is given in terms of the Pauli matrices $(\sigma_x,\sigma_y)$ as
 \begin{equation}
 \label{eq:GeneralRot}
  \hat U(t) = {\rm exp}\left(\frac{i \Omega_R t}{2}\left[\cos(\varphi)\hat\sigma_x - \sin(\varphi)\hat\sigma_y\right]\right)
 \end{equation} 
 where $\Omega_R=\lambda |\bra{\psi_1}\hat V\ket{\psi_2}|/\hbar$ is the Rabi frequency~\cite{cold:scul02}.
 This enables arbitrary rotations about any axis in the $x y$ plane of the 
 system's Bloch sphere. A $z$ rotation may be driven by combining rotations about the 
 $x$ and $y$ axes.

 Driving a single-qubit operation at a certain lattice site necessitates that the 
 energy difference between $\ket{0}$ and $\ket{1}$ differs from that of other lattice sites. 
 This can be achieved by shining a laser with a waist on the order of the lattice 
 spacing perpendicular to the lattice, as was accomplished experimentally in~\cite{cold:weit11}. 
 The perturbation by this laser can be assumed to have the Gaussian form 
 $V_{\rm Gauss}=-\gamma \, {\rm exp}[{(x-x_0)^2/(2\sigma^2)}]$. 
 Depending on whether the atoms are strong or weak-field seekers one has $\gamma>0$
 or $\gamma<0$ and the additional intensity will enlarge or reduce the energy difference 
 $\hbar(\omega_2-\omega_1)$ between ground and excited state [see Fig.~\ref{fig:system} c)]. 
 By shaking with a frequency $\omega_{\rm sh}=\omega_2-\omega_1$ one is then able to 
 drive single-qubit rotations on the marked lattice site.
 
 The detuning of the energy levels leads to an additional rotation about 
 the $z$ axis in the Bloch sphere of the marked and also of neighboring qubits.
 Moreover, off-resonant shaking of a lattice site can also induce a slight
 phase shift~\cite{cold:vand05} leading to a general dephasing of the qubits.
 These effects lead predominantly to additional terms in the Hamiltonian of the form 
 $\hat W = \sum_j a_j\, \hat \sigma_z^{(j)}$, 
 where the Pauli $z$ operator $\sigma_z^{(j)}$ acts on qubit $j$. 
 If the $a_i$'s are known one can account for the additional $z$ rotations
 within the quantum calculation. 
 However, one can also cancel these rotations by applying a scheme similar to the 
 refocussing technique well known from NMR quantum control \cite{cold:vand05}. 
 Suppose an $x$ rotation $\hat X_{\phi}^{(s)}={\rm exp}(i\phi \hat\sigma_x^{(s)}/2 )$ 
 is to be driven on a qubit at site $s$. The time $2\tau$ of the operation is chosen such that
 ${\rm exp}(- i \tau a_s \hat\sigma_z^{(s)}/2 )=1$.
 Then, inserting a global $\pi$ rotation $\hat X_{\pi}^{\rm gl} =
 \exp(i \pi/2  \sum_j \hat \sigma_x^{(j)} )$ before and in the middle of the operation 
 does not perturb the $\hat X_{\phi}^{(s)}$ rotation but all $z$ rotations are cancelled
 since $\hat X_{\pi}^{\rm gl}$ has the property that 
 $\hat X_{\pi}^{\rm gl} e^{-i t\hat W /\hbar} \hat X_{\pi}^{\rm gl} e^{-i t\hat W /\hbar} = 1$.
 
\begin{figure}[htp]
 \centering
 \includegraphics[width=1.0\linewidth]{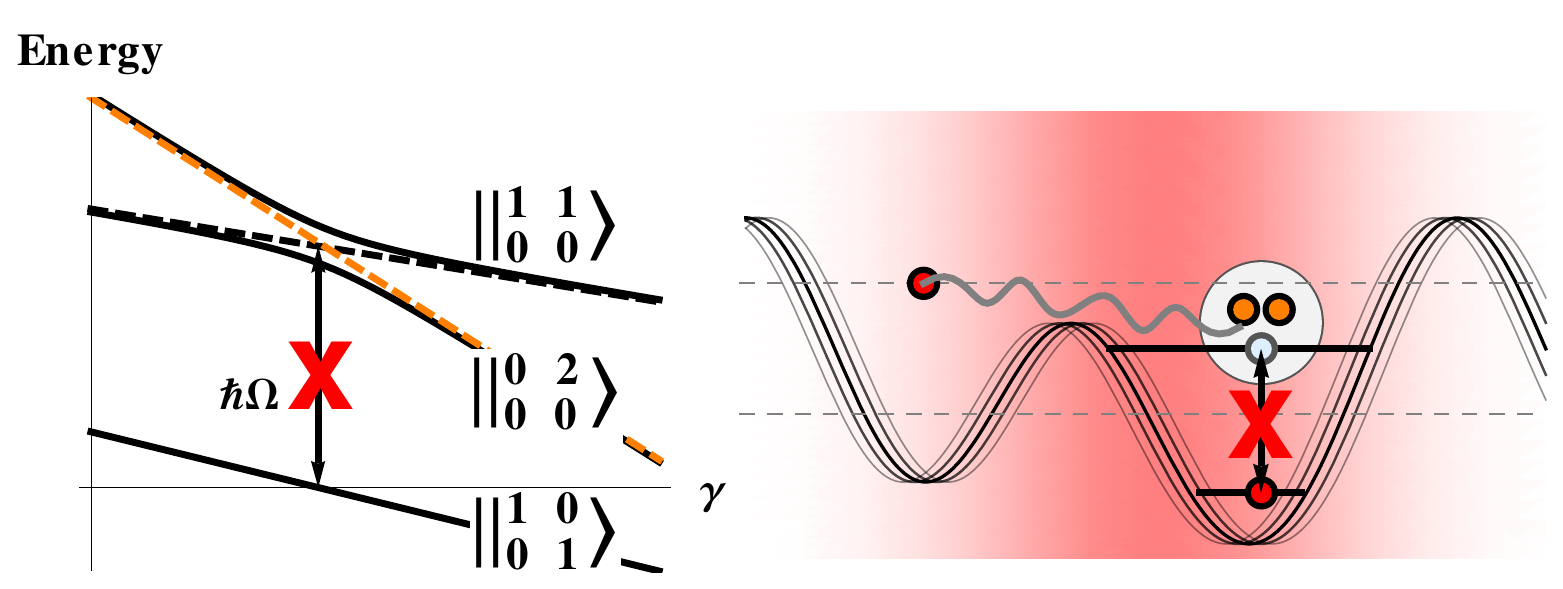}
 \caption{(Color online) 
  A laser (red shading in right image) is positioned slightly offset from the middle between two
  lattice sites so that both lattice sites are coupled and the energy of the right lattice site is
  lower than the one of the left site. At a certain laser intensity the energy of the repulsively bound
  state $\state{0}{0}{0}{2}$ and the state $\state{0}{1}{0}{1}$ form an avoided crossing 
  (left image). It is now possible to shake the lattice with frequency $\Omega$ resonantly 
  to the transition $\ket{1 0}\leftrightarrow\ket{1 1}$ 
  ($\state{1}{0}{0}{1}\leftrightarrow\state{1}{0}{1}{0}$) 
  while the transition $\ket{0 0}\leftrightarrow\ket{0 1}$
  ($\state{0}{1}{0}{1}\leftrightarrow\state{0}{1}{1}{0}$) is off-resonant. 
  This enables rotations on the right target qubit conditioned by the left control qubit.
}
 \label{fig:CROTScheme}
\end{figure}

 Together with single-qubit operations, the ability to drive a controlled rotation (CROT) 
 between adjacent lattice sites completes a universal gate set \cite{cold:brem02}. 
 A CROT rotates one qubit (the target qubit) if and only if another qubit 
 (the control qubit) is in state $\ket{1}$.
 The strategy to perform this operation between neighboring qubits is to deform the 
 lattice such that a repulsively bound state $\state{0}{0}{0}{2}$
 \cite{cold:wink06} comes into resonance with the state $\ket{0 0}=\state{0}{1}{0}{1}$. 
 The coupling between the lattice sites leads to an avoided crossing in the energy spectrum 
 (see Fig.~\ref{fig:CROTScheme}).
 If we identify the left qubit with the control qubit, any rotation like that of 
 Eq.~(\ref{eq:GeneralRot}) on the right target qubit becomes 
 off-resonant and is inhibited iff the control qubit is initially 
 in the excited state ($\ket{0}$).
 
 In order to validate the proposal for single and two-qubit operations by a numerical 
 study a lattice with unit filling described by a Wannier basis of the first three Bloch 
 bands with periodic boundary conditions is considered. It is assumed that the lattice 
 is sufficiently deep so that, as usual within the Bose-Hubbard model, only next-neighbor 
 hopping and onsite interaction need to be considered. 
 For the single-qubit rotation, a system of two lattice sites
 suffices to estimate the influence of the operation on the remaining system 
 and vice versa. For studying the two-qubit operation a third site
 has to be added.
 The third Bloch band is included to study possible excitations of
 atoms out of the qubit basis.
 In the following a lattice depth  $V_0 = 2.7 \hbar\omega_L = 29.16$ is considered.
 For this lattice an interaction strength $g=1.87$ suffices to form a Mott insulator
 so that all qubit states are eigenstates of the Hamiltonian.

 As an example for a single-qubit operation the NOT operation on the right site 
 of a two-well lattice is considered. As stated above the dephasing caused by terms 
 of the form $\hat W = a_1 \hat \sigma_z^{(1)} + a_2 \hat \sigma_z^{(2)}$
 is inhibited by two global $\hat X_{\pi}^{\rm gl}$ rotations.
 However, the narrow-waist laser can also lead to a weak coupling 
 $b \hat \sigma_z^{(1)} \cdot \hat \sigma_z^{(2)}$ which can only 
 be cancelled by more complex sequences of refocussing pulses. To suppress this 
 term during the operation the lattice depth $V_0$ is temporarily enlarged 
 to $V_0 +\delta V_0$ before shining in the 
 narrow-waist laser.
 Both perturbations 
 are switched on and  off adiabatically and are sufficiently small so that
 couplings to bands above the third Bloch band are negligible.
 While the perturbations are active $\hat V_{\rm sh}$ drives resonantly in the adiabatic basis
 the rotation $\hat U = {\rm exp}(i \Omega_R t \hat \sigma_x/2)$ on the right site, 
 which for $\Omega_R t=\pi$ is the NOT operation up to a global phase. 
\begin{figure}[htp]
 \centering
 \includegraphics[width=0.9\linewidth]{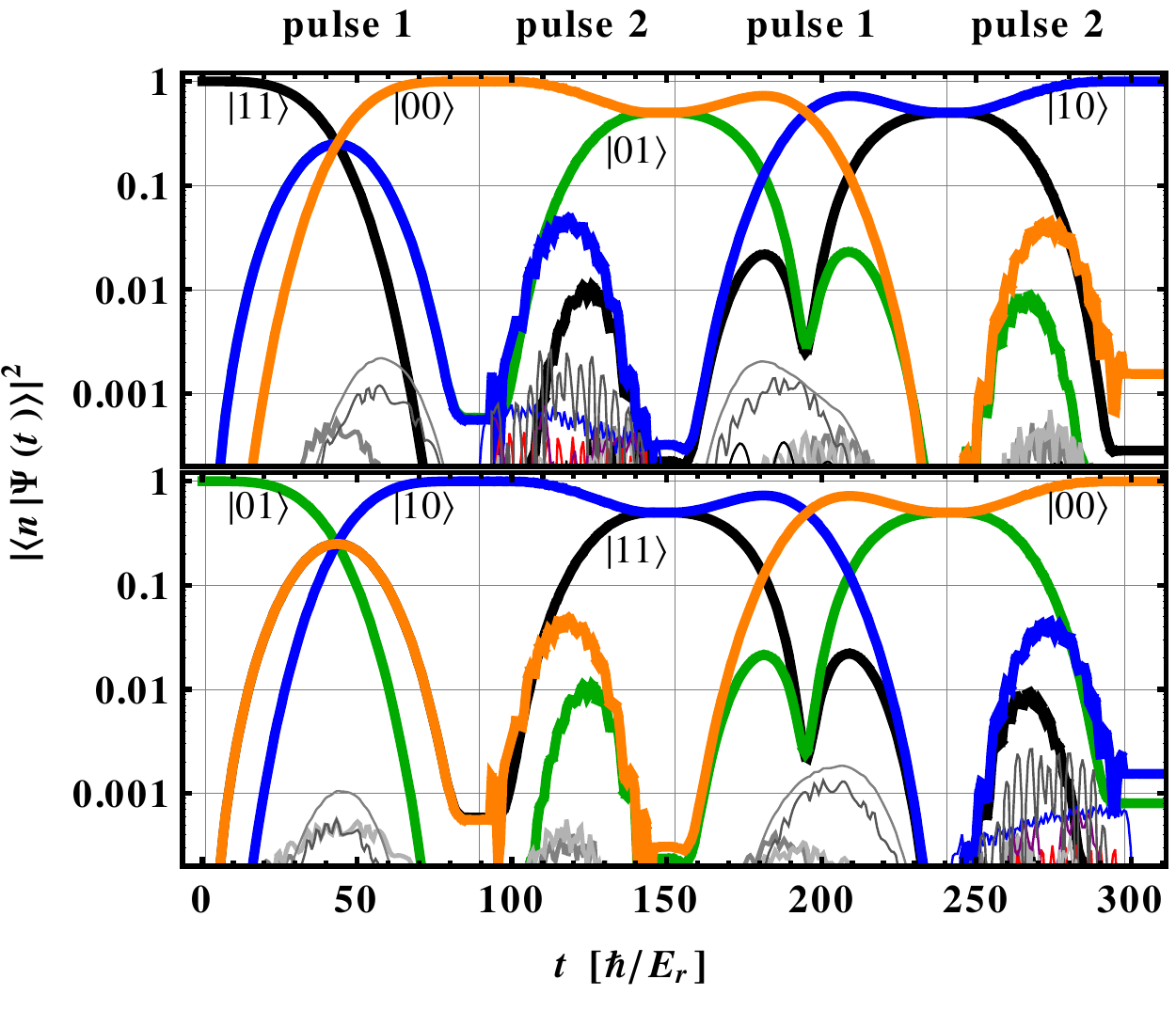}
 \caption{(Color online)
 Admixtures of different eigenstates for the initial states $\ket{11}$ and $\ket{01}$ 
 during a NOT operation on the right qubit of a two-well lattice.
 Pulse 1 drives an $\hat X_{\pi}^{\rm gl}$ rotation. 
 Pulse 2 starts by linearly increasing the lattice depth
 by $\delta V_0 = 0.17 V_0$ during a ramping time of $5.9 \hbar/E_r$.
 Then the Gaussian beam with waist $\sigma=\pi/2$ centered at the site of the target qubit 
 is linearly ramped during a time of $0.4 \hbar/E_r$ to the strength $\gamma=2.62$.
 After a $\pi/2$ rotation ($\sqrt{\rm NOT}$) on the right qubit driven by shaking the 
 lattice resonantly to the energy difference between $\ket{11}$ and $\ket{10}$, 
 $\delta V_0$ and $\gamma$ are ramped off in reversed order.
 The simulations show a fidelity of 99.7\% for a gate time of $300\hbar/E_r$.
}
 \label{fig:NOT}
\end{figure}
 Fig.~\ref{fig:NOT} shows numerical simulations for the NOT operation including
 two $\hat X_{\phi}^{\rm gl}$ operations for refocussing. 
 For the chosen parameter set (see caption of Fig.~\ref{fig:NOT}) the 
 influence of the qubit state on the left side is negligible 
 and an average fidelity~\cite{cold:niel02} of 99.7\% is reached.
 Without refocussing (not shown) the fidelity would be only 33.2\% due to dephasing.

\begin{figure}[htp]
 \centering
 \includegraphics[width=0.9\linewidth]{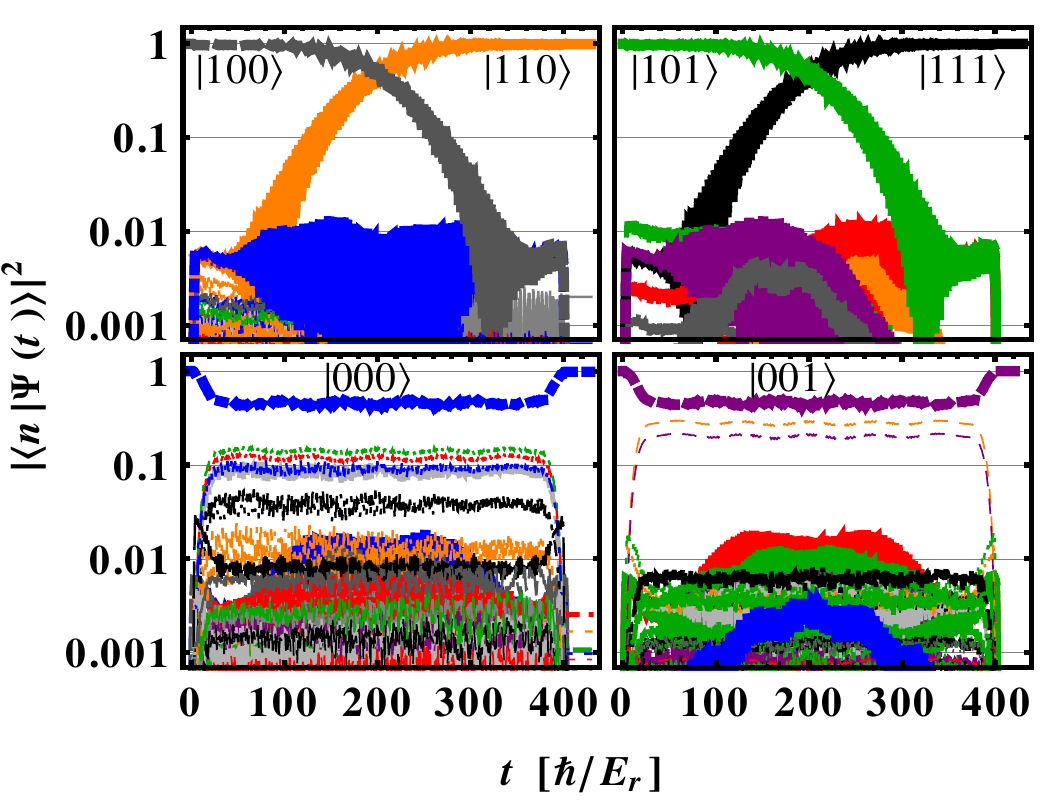}
 \caption{(Color online)
 Admixtures of different eigenstates for different initial states during a CROT 
 on the left control qubit at $x=0$ and the central target qubit at $x=\pi$. 
 The Gaussian beam with waist $\sigma=\pi/2$ at $x_0=0.6 \pi$ is linearly ramped to
 a strength of $\gamma=0.224$ during a time of $2\hbar/E_r$. After waiting for $12\hbar/E_r$
 the beam is linearly ramped for $16\hbar/E_r$ to the avoided crossing
 with the repulsively bound state at $\gamma=0.204$.
 After shaking the lattice resonantly to the energy difference between
 $\ket{111}$ and $\ket{101}$ the gaussian beam is switched off in reversed order.
 For a gate time of $400\hbar/E_r$ a fidelity of 99.4\% is reached (see text).
}
 \label{fig:CROT}
\end{figure}

 For the CROT operation a system of three lattice sites is considered. 
 The left site acts as the control qubit of an $\hat X_{\pi}^{(2)}$ rotation on
 the central target qubit. 
 Up to a local phase this operation is equivalent to a controlled-NOT operation.
 Due to the periodic boundary conditions the right site is coupled to the 
 control and the target qubit.
 Example results and parameters of the numerical simulations of the CROT 
 are shown in Fig.~\ref{fig:CROT}.
 For a gate time of $400 \hbar/E_r$ an average fidelity of 99.4\% is reached.
 For the evaluation of the fidelity the individual $z$ rotations and those $z$ rotations
 coupled to the control qubit 
 are neglected. This is possible since equipped with the global $\hat X_\pi^{\rm gl}$ and local NOT 
 operations these $z$ rotation can be easily cancelled by using a refocusing scheme.
 Opposing to the NOT operation an important source of infidelity of the CROT stems 
 from the leakage to states out of the qubit basis. Averaged over the computational basis 
 the leakage probability after the operation is 0.27\%.
 To diminish this one can either use techniques of leakage elimination \cite{cold:byrd05}
 or one has to choose a deeper lattice.
 This leads to weaker coupling between the lattice sites and thus to longer gate times.
 Increasing, e.g., the lattice depth $V_0$ from $2.7\hbar\omega_L$ to $2.8\hbar\omega_L$
 and the gate time to $540\hbar/E_r$ reduces the leakage probability to 0.15\% and 
 increases the fidelity by another 0.06\%.

 Independently of the chosen lattice depth, the time scale $\hbar/E_r$ is a crucial 
 system parameter for the speed and accuracy of the manipulations. 
 In general a large recoil energy $E_r=\hbar^2k^2/ 2m$ allows shorter time scales. 
 For example, the NOT operation for a $^{87}$Rb
 system in a $d=500\,$nm OL ($E_r=1.5\times 10^{-30}\,$J) lasts $21\,$ms while the 
 same operation would take $0.6\,$ms for $^{7}$Li in a $d=300\,$nm lattice 
 ($E_r=5.2\times 10^{-29}\,$J).

 In order for the operations to be robust any relative energy shift of the manipulated 
 qubit states by some external perturbation must be small compared to the energy scale
 of the Rabi oscillations $\hbar \Omega_R$ which in our case is about $0.1\,E_r$.
 Since the energy differences are mainly influenced by the lattice potential itself,
 its uncontrolled perturbation can severely reduce the fidelity of the operation. 
 Considering the CROT operation, which is more sensitive to perturbations than 
 single-qubit operations, only a lattice laser intensity which is controlled on the 
 $10^{-4}$ level leads to negligible fidelity reduction (see Tab.~\ref{tab:fidelities}). 
 On the other hand, perturbations of other parameters such as the intensity of the 
 gaussian laser beam or the interaction strength are much less severe.

\begin{table}[t]
  \caption{  \label{tab:fidelities}
  Average fidelity of the CROT operation presented in Fig.~\ref{fig:CROT} for uncontrolled errors 
  of the interaction strength $g$, the strength of the gaussian laser $\gamma$, and the
  lattice depth $V_0$.}
\begin{tabular}{@{}l|c|c|c|c|c|c|c}
\hline 
\hline 
  Parameter &\multicolumn{2}{c|}{$g$} & \multicolumn{2}{c}{$\gamma$} & \multicolumn{3}{|c}{$V_0$}\\
\hline
  Error     & 0.1\% & 1\% & 0.1\% & 1\% & 0.01\% & 0.1\% & 1\% \\

\hline
Fidelity &
99.4\% & 98.1\% &
98.9\% & 97.3\% &
99.2\% & 90.9\% & 31.4\% \\
\hline
\hline
      \end{tabular}
\end{table}

\begin{figure}[htp]
 \centering
 \includegraphics[width=0.9\linewidth]{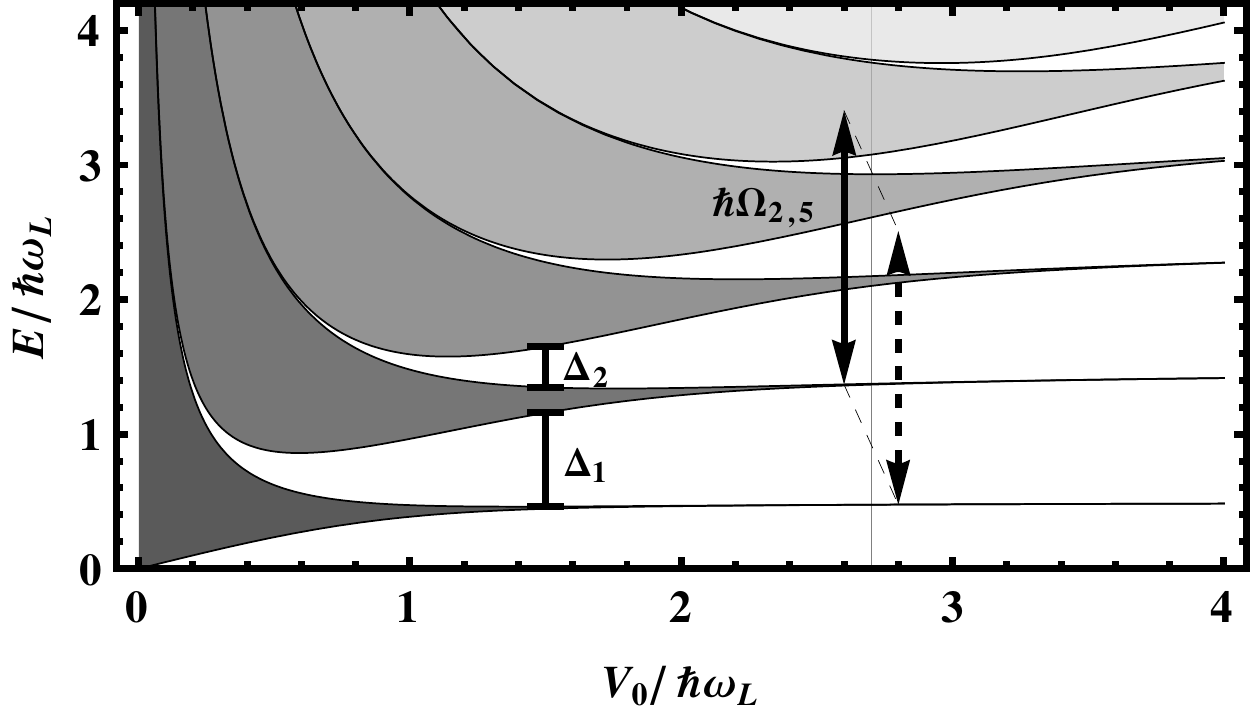}
 \caption{
 Energy widths of the Bloch bands as a function of the lattice depth. 
 For shallow lattices the band gap $\Delta_{2}$ is much smaller than
 $\Delta_{1}$. When moving the lattice this allows for Landau-Zener transitions 
 out of the $2^{\rm nd}$ band while atoms in the $1^{\rm st}$ band are dragged
 by the lattice.
 At $V_0=2.7\hbar\omega$ one may drive a transition from the $2^{\rm nd}$ to the
 $5^{\rm th}$ Bloch band with energy difference $\hbar \Omega_{2,5}$ while a transition from
 the first Bloch band is forbidden by a band gap.}
 \label{fig:bloch_bands}
\end{figure}

 The last important ingredient for quantum computation is the possibility to read out qubit 
 states with high fidelity.
 This can be done by removing atoms in the excited state $\ket{0}$ from the lattice
 and determining subsequently the atom distribution of the remaining atoms
 by fluorescence imaging with single-site resolution. 
 The removal of excited atoms is related to an evaporative cooling of the system 
 and corresponding strategies may be applied. One method is to accelerate the lattice in $x$ direction.
 For sufficiently deep lattices the large gap between the first and second Bloch bands inhibits 
 Landau-Zener transitions such that atoms in the ground state $\ket{1}$ are dragged by the lattice. 
 However, atoms in state $\ket{0}$ may tunnel to higher Bloch bands and eventually leave the
 lattice~\cite{cold:niu96,cold:madi98}.
 The tunneling can be enhanced by first transferring atoms in state $\ket{0}$ to a higher 
 Bloch band. 
 As shown in Fig.~\ref{fig:bloch_bands} for the exemplary lattice depth of
 $V_0 = 2.7\hbar\omega_{L}$ one can drive transitions from the 2$^{\rm nd}$ to the 
 high-lying 5$^{\rm th}$ Bloch band by means of shaking the lattice with frequency $\Omega_{2,5}$ while 
 transitions from the first Bloch band are inhibited due to a band gap.
 In the 5th Bloch band the atoms are quasi-free and leave the lattice even 
 during slow accelerations.
 Lately, a closely related scheme of a state-dependent removal of atoms from an optical lattice
 has been used experimentally for cooling a quantum gas \cite{cold:bakr11} supporting  
 the practicability of vibrationally encoded qubits and the proposed read-out scheme.

 In conclusion, a scheme for quantum computation with Bosons in 1d 
 optical lattices was presented.
 The qubits are encoded in the spacial atomic wave function which suppresses decoherence due to
 fluctuating magnetic fields.
 It was shown that by shaking the lattice one may drive single and controlled qubit rotations 
 and how qubit dephasing can be prevented by using refocussing pulses.
 The qubit readout can be performed by removing atoms in excited states from the lattice
 and determining the atom distribution by fluorescence imaging. 
 For gate times on the order of milliseconds fidelities above 99\% can be 
 reached. We believe that gate times and fidelities can be further improved, e.g., by using 
 optimal control.
 It would be interesting to extend the approach to a 2d OL in order to reduce 
 the number of needed swap operations. However, considering $^{7}$Li atoms in a $d=300\,$nm 
 lattice, already within the 1d approach a factorization of 15 would be 
 feasible within about 50 ms with about 20 ms needed for swapping operations.

 \acknowledgments
 We thank F. Schmidt-Kaler, J. Simon, and S. Kuhr for helpful comments. 
  This work was supported by the {\it Deutsche Telekom Stiftung} and the 
 {\it Fonds der Chemischen Industrie}.


\begin{thebibliography}{29}%
\makeatletter
\providecommand \@ifxundefined [1]{%
 \@ifx{#1\undefined}
}%
\providecommand \@ifnum [1]{%
 \ifnum #1\expandafter \@firstoftwo
 \else \expandafter \@secondoftwo
 \fi
}%
\providecommand \@ifx [1]{%
 \ifx #1\expandafter \@firstoftwo
 \else \expandafter \@secondoftwo
 \fi
}%
\providecommand \natexlab [1]{#1}%
\providecommand \enquote  [1]{``#1''}%
\providecommand \bibnamefont  [1]{#1}%
\providecommand \bibfnamefont [1]{#1}%
\providecommand \citenamefont [1]{#1}%
\providecommand \href@noop [0]{\@secondoftwo}%
\providecommand \href [0]{\begingroup \@sanitize@url \@href}%
\providecommand \@href[1]{\@@startlink{#1}\@@href}%
\providecommand \@@href[1]{\endgroup#1\@@endlink}%
\providecommand \@sanitize@url [0]{\catcode `\\12\catcode `\$12\catcode
  `\&12\catcode `\#12\catcode `\^12\catcode `\_12\catcode `\%12\relax}%
\providecommand \@@startlink[1]{}%
\providecommand \@@endlink[0]{}%
\providecommand \url  [0]{\begingroup\@sanitize@url \@url }%
\providecommand \@url [1]{\endgroup\@href {#1}{\urlprefix }}%
\providecommand \urlprefix  [0]{URL }%
\providecommand \Eprint [0]{\href }%
\providecommand \doibase [0]{http://dx.doi.org/}%
\providecommand \selectlanguage [0]{\@gobble}%
\providecommand \bibinfo  [0]{\@secondoftwo}%
\providecommand \bibfield  [0]{\@secondoftwo}%
\providecommand \translation [1]{[#1]}%
\providecommand \BibitemOpen [0]{}%
\providecommand \bibitemStop [0]{}%
\providecommand \bibitemNoStop [0]{.\EOS\space}%
\providecommand \EOS [0]{\spacefactor3000\relax}%
\providecommand \BibitemShut  [1]{\csname bibitem#1\endcsname}%
\let\auto@bib@innerbib\@empty
\bibitem [{\citenamefont {Bloch}(2008)}]{cold:bloc08a}%
  \BibitemOpen
  \bibfield  {author} {\bibinfo {author} {\bibfnamefont {I.}~\bibnamefont
  {Bloch}},\ }\href@noop {} {\bibfield  {journal} {\bibinfo  {journal}
  {Nature}\ }\textbf {\bibinfo {volume} {453}},\ \bibinfo {pages} {1016}
  (\bibinfo {year} {2008})}\BibitemShut {NoStop}%
\bibitem [{\citenamefont {Schrader}\ \emph {et~al.}(2004)\citenamefont
  {Schrader}, \citenamefont {Dotsenko}, \citenamefont {Khudaverdyan},
  \citenamefont {Miroshnychenko}, \citenamefont {Rauschenbeutel},\ and\
  \citenamefont {Meschede}}]{cold:schr04}%
  \BibitemOpen
  \bibfield  {author} {\bibinfo {author} {\bibfnamefont {D.}~\bibnamefont
  {Schrader}}, \bibinfo {author} {\bibfnamefont {I.}~\bibnamefont {Dotsenko}},
  \bibinfo {author} {\bibfnamefont {M.}~\bibnamefont {Khudaverdyan}}, \bibinfo
  {author} {\bibfnamefont {Y.}~\bibnamefont {Miroshnychenko}}, \bibinfo
  {author} {\bibfnamefont {A.}~\bibnamefont {Rauschenbeutel}}, \ and\ \bibinfo
  {author} {\bibfnamefont {D.}~\bibnamefont {Meschede}},\ }\href@noop {}
  {\bibfield  {journal} {\bibinfo  {journal} {Phys.\,Rev.\,Lett.}\ }\textbf
  {\bibinfo {volume} {93}},\ \bibinfo {pages} {150501} (\bibinfo {year}
  {2004})}\BibitemShut {NoStop}%
\bibitem [{\citenamefont {Karski}\ \emph {et~al.}(2009)\citenamefont {Karski},
  \citenamefont {F\"orster}, \citenamefont {Choi}, \citenamefont {Alt},
  \citenamefont {Widera},\ and\ \citenamefont {Meschede}}]{cold:kars09}%
  \BibitemOpen
  \bibfield  {author} {\bibinfo {author} {\bibfnamefont {M.}~\bibnamefont
  {Karski}}, \bibinfo {author} {\bibfnamefont {L.}~\bibnamefont {F\"orster}},
  \bibinfo {author} {\bibfnamefont {J.~M.}\ \bibnamefont {Choi}}, \bibinfo
  {author} {\bibfnamefont {W.}~\bibnamefont {Alt}}, \bibinfo {author}
  {\bibfnamefont {A.}~\bibnamefont {Widera}}, \ and\ \bibinfo {author}
  {\bibfnamefont {D.}~\bibnamefont {Meschede}},\ }\href@noop {} {\bibfield
  {journal} {\bibinfo  {journal} {Phys.\,Rev.\,Lett.}\ }\textbf {\bibinfo
  {volume} {102}},\ \bibinfo {pages} {053001} (\bibinfo {year}
  {2009})}\BibitemShut {NoStop}%
\bibitem [{\citenamefont {Bakr}\ \emph {et~al.}(2009)\citenamefont {Bakr},
  \citenamefont {Gillen}, \citenamefont {Peng}, \citenamefont {Folling},\ and\
  \citenamefont {Greiner}}]{cold:bakr09}%
  \BibitemOpen
  \bibfield  {author} {\bibinfo {author} {\bibfnamefont {W.~S.}\ \bibnamefont
  {Bakr}}, \bibinfo {author} {\bibfnamefont {J.~I.}\ \bibnamefont {Gillen}},
  \bibinfo {author} {\bibfnamefont {A.}~\bibnamefont {Peng}}, \bibinfo {author}
  {\bibfnamefont {S.}~\bibnamefont {Folling}}, \ and\ \bibinfo {author}
  {\bibfnamefont {M.}~\bibnamefont {Greiner}},\ }\href@noop {} {\bibfield
  {journal} {\bibinfo  {journal} {Nature}\ }\textbf {\bibinfo {volume} {462}},\
  \bibinfo {pages} {74} (\bibinfo {year} {2009})}\BibitemShut {NoStop}%
\bibitem [{\citenamefont {Sherson}\ \emph {et~al.}(2010)\citenamefont
  {Sherson}, \citenamefont {Weitenberg}, \citenamefont {Endres}, \citenamefont
  {Cheneau}, \citenamefont {Bloch},\ and\ \citenamefont {Kuhr}}]{cold:sher10}%
  \BibitemOpen
  \bibfield  {author} {\bibinfo {author} {\bibfnamefont {J.~F.}\ \bibnamefont
  {Sherson}}, \bibinfo {author} {\bibfnamefont {C.}~\bibnamefont {Weitenberg}},
  \bibinfo {author} {\bibfnamefont {M.}~\bibnamefont {Endres}}, \bibinfo
  {author} {\bibfnamefont {M.}~\bibnamefont {Cheneau}}, \bibinfo {author}
  {\bibfnamefont {I.}~\bibnamefont {Bloch}}, \ and\ \bibinfo {author}
  {\bibfnamefont {S.}~\bibnamefont {Kuhr}},\ }\href@noop {} {\bibfield
  {journal} {\bibinfo  {journal} {Nature}\ }\textbf {\bibinfo {volume} {467}},\
  \bibinfo {pages} {68} (\bibinfo {year} {2010})}\BibitemShut {NoStop}%
\bibitem [{\citenamefont {Bakr}\ \emph {et~al.}(2010)\citenamefont {Bakr},
  \citenamefont {Peng}, \citenamefont {Tai}, \citenamefont {Ma}, \citenamefont
  {Simon}, \citenamefont {Gillen}, \citenamefont {F\"olling}, \citenamefont
  {Pollet},\ and\ \citenamefont {Greiner}}]{cold:bakr10}%
  \BibitemOpen
  \bibfield  {author} {\bibinfo {author} {\bibfnamefont {W.~S.}\ \bibnamefont
  {Bakr}}, \bibinfo {author} {\bibfnamefont {A.}~\bibnamefont {Peng}}, \bibinfo
  {author} {\bibfnamefont {M.~E.}\ \bibnamefont {Tai}}, \bibinfo {author}
  {\bibfnamefont {R.}~\bibnamefont {Ma}}, \bibinfo {author} {\bibfnamefont
  {J.}~\bibnamefont {Simon}}, \bibinfo {author} {\bibfnamefont {J.~I.}\
  \bibnamefont {Gillen}}, \bibinfo {author} {\bibfnamefont {S.}~\bibnamefont
  {F\"olling}}, \bibinfo {author} {\bibfnamefont {L.}~\bibnamefont {Pollet}}, \
  and\ \bibinfo {author} {\bibfnamefont {M.}~\bibnamefont {Greiner}},\
  }\href@noop {} {\bibfield  {journal} {\bibinfo  {journal} {Science}\ }\textbf
  {\bibinfo {volume} {329}},\ \bibinfo {pages} {547} (\bibinfo {year}
  {2010})}\BibitemShut {NoStop}%
\bibitem [{\citenamefont {Weitenberg}\ \emph {et~al.}(2011)\citenamefont
  {Weitenberg}, \citenamefont {Endres}, \citenamefont {Sherson}, \citenamefont
  {Cheneau}, \citenamefont {Schausz}, \citenamefont {Fukuhara}, \citenamefont
  {Bloch},\ and\ \citenamefont {Kuhr}}]{cold:weit11}%
  \BibitemOpen
  \bibfield  {author} {\bibinfo {author} {\bibfnamefont {C.}~\bibnamefont
  {Weitenberg}}, \bibinfo {author} {\bibfnamefont {M.}~\bibnamefont {Endres}},
  \bibinfo {author} {\bibfnamefont {J.~F.}\ \bibnamefont {Sherson}}, \bibinfo
  {author} {\bibfnamefont {M.}~\bibnamefont {Cheneau}}, \bibinfo {author}
  {\bibfnamefont {P.}~\bibnamefont {Schausz}}, \bibinfo {author} {\bibfnamefont
  {T.}~\bibnamefont {Fukuhara}}, \bibinfo {author} {\bibfnamefont
  {I.}~\bibnamefont {Bloch}}, \ and\ \bibinfo {author} {\bibfnamefont
  {S.}~\bibnamefont {Kuhr}},\ }\href@noop {} {\bibfield  {journal} {\bibinfo
  {journal} {Nature}\ }\textbf {\bibinfo {volume} {471}},\ \bibinfo {pages}
  {319} (\bibinfo {year} {2011})}\BibitemShut {NoStop}%
\bibitem [{\citenamefont {Ladd}\ \emph {et~al.}(2010)\citenamefont {Ladd},
  \citenamefont {Jelezko}, \citenamefont {Laflamme}, \citenamefont {Nakamura},
  \citenamefont {Monroe},\ and\ \citenamefont {O'Brien}}]{cold:ladd10}%
  \BibitemOpen
  \bibfield  {author} {\bibinfo {author} {\bibfnamefont {T.~D.}\ \bibnamefont
  {Ladd}}, \bibinfo {author} {\bibfnamefont {F.}~\bibnamefont {Jelezko}},
  \bibinfo {author} {\bibfnamefont {R.}~\bibnamefont {Laflamme}}, \bibinfo
  {author} {\bibfnamefont {Y.}~\bibnamefont {Nakamura}}, \bibinfo {author}
  {\bibfnamefont {C.}~\bibnamefont {Monroe}}, \ and\ \bibinfo {author}
  {\bibfnamefont {J.~L.}\ \bibnamefont {O'Brien}},\ }\href@noop {} {\bibfield
  {journal} {\bibinfo  {journal} {Nature}\ }\textbf {\bibinfo {volume} {464}},\
  \bibinfo {pages} {45} (\bibinfo {year} {2010})}\BibitemShut {NoStop}%
\bibitem [{\citenamefont {Brennen}\ \emph {et~al.}(1999)\citenamefont
  {Brennen}, \citenamefont {Caves}, \citenamefont {Jessen},\ and\ \citenamefont
  {Deutsch}}]{cold:bren99}%
  \BibitemOpen
  \bibfield  {author} {\bibinfo {author} {\bibfnamefont {G.~K.}\ \bibnamefont
  {Brennen}}, \bibinfo {author} {\bibfnamefont {C.~M.}\ \bibnamefont {Caves}},
  \bibinfo {author} {\bibfnamefont {P.~S.}\ \bibnamefont {Jessen}}, \ and\
  \bibinfo {author} {\bibfnamefont {I.~H.}\ \bibnamefont {Deutsch}},\
  }\href@noop {} {\bibfield  {journal} {\bibinfo  {journal}
  {Phys.\,Rev.\,Lett.}\ }\textbf {\bibinfo {volume} {82}},\ \bibinfo {pages}
  {1060} (\bibinfo {year} {1999})}\BibitemShut {NoStop}%
\bibitem [{\citenamefont {Hayes}\ \emph {et~al.}(2007)\citenamefont {Hayes},
  \citenamefont {Julienne},\ and\ \citenamefont {Deutsch}}]{cold:haye07}%
  \BibitemOpen
  \bibfield  {author} {\bibinfo {author} {\bibfnamefont {D.}~\bibnamefont
  {Hayes}}, \bibinfo {author} {\bibfnamefont {P.~S.}\ \bibnamefont {Julienne}},
  \ and\ \bibinfo {author} {\bibfnamefont {I.~H.}\ \bibnamefont {Deutsch}},\
  }\href@noop {} {\bibfield  {journal} {\bibinfo  {journal}
  {Phys.\,Rev.\,Lett.}\ }\textbf {\bibinfo {volume} {98}},\ \bibinfo {pages}
  {070501} (\bibinfo {year} {2007})}\BibitemShut {NoStop}%
\bibitem [{\citenamefont {Daley}\ \emph {et~al.}(2008)\citenamefont {Daley},
  \citenamefont {Boyd}, \citenamefont {Ye},\ and\ \citenamefont
  {Zoller}}]{cold:dale08}%
  \BibitemOpen
  \bibfield  {author} {\bibinfo {author} {\bibfnamefont {A.~J.}\ \bibnamefont
  {Daley}}, \bibinfo {author} {\bibfnamefont {M.~M.}\ \bibnamefont {Boyd}},
  \bibinfo {author} {\bibfnamefont {J.}~\bibnamefont {Ye}}, \ and\ \bibinfo
  {author} {\bibfnamefont {P.}~\bibnamefont {Zoller}},\ }\href@noop {}
  {\bibfield  {journal} {\bibinfo  {journal} {Phys.\,Rev.\,Lett.}\ }\textbf
  {\bibinfo {volume} {101}},\ \bibinfo {pages} {170504} (\bibinfo {year}
  {2008})}\BibitemShut {NoStop}%
\bibitem [{\citenamefont {Negretti}\ \emph {et~al.}(2011)\citenamefont
  {Negretti}, \citenamefont {Treutlein},\ and\ \citenamefont
  {Calarco}}]{cold:negr11}%
  \BibitemOpen
  \bibfield  {author} {\bibinfo {author} {\bibfnamefont {A.}~\bibnamefont
  {Negretti}}, \bibinfo {author} {\bibfnamefont {P.}~\bibnamefont {Treutlein}},
  \ and\ \bibinfo {author} {\bibfnamefont {T.}~\bibnamefont {Calarco}},\
  }\href@noop {} {\bibfield  {journal} {\bibinfo  {journal} {Quantum
  Information Processing}\ }\textbf {\bibinfo {volume} {10}},\ \bibinfo {pages}
  {721} (\bibinfo {year} {2011})}\BibitemShut {NoStop}%
\bibitem [{\citenamefont {Anderlini}\ \emph {et~al.}(2007)\citenamefont
  {Anderlini}, \citenamefont {Lee}, \citenamefont {Brown}, \citenamefont
  {Sebby-Strabley}, \citenamefont {Phillips},\ and\ \citenamefont
  {Porto}}]{cold:ande07}%
  \BibitemOpen
  \bibfield  {author} {\bibinfo {author} {\bibfnamefont {M.}~\bibnamefont
  {Anderlini}}, \bibinfo {author} {\bibfnamefont {P.~J.}\ \bibnamefont {Lee}},
  \bibinfo {author} {\bibfnamefont {B.~L.}\ \bibnamefont {Brown}}, \bibinfo
  {author} {\bibfnamefont {J.}~\bibnamefont {Sebby-Strabley}}, \bibinfo
  {author} {\bibfnamefont {W.~D.}\ \bibnamefont {Phillips}}, \ and\ \bibinfo
  {author} {\bibfnamefont {J.~V.}\ \bibnamefont {Porto}},\ }\href@noop {}
  {\bibfield  {journal} {\bibinfo  {journal} {Nature}\ }\textbf {\bibinfo
  {volume} {448}},\ \bibinfo {pages} {452} (\bibinfo {year}
  {2007})}\BibitemShut {NoStop}%
\bibitem [{\citenamefont {Eckert}\ \emph {et~al.}(2002)\citenamefont {Eckert},
  \citenamefont {Mompart}, \citenamefont {Yi}, \citenamefont {Schliemann},
  \citenamefont {Bru\ss{}}, \citenamefont {Birkl},\ and\ \citenamefont
  {Lewenstein}}]{cold:ecke02}%
  \BibitemOpen
  \bibfield  {author} {\bibinfo {author} {\bibfnamefont {K.}~\bibnamefont
  {Eckert}}, \bibinfo {author} {\bibfnamefont {J.}~\bibnamefont {Mompart}},
  \bibinfo {author} {\bibfnamefont {X.~X.}\ \bibnamefont {Yi}}, \bibinfo
  {author} {\bibfnamefont {J.}~\bibnamefont {Schliemann}}, \bibinfo {author}
  {\bibfnamefont {D.}~\bibnamefont {Bru\ss{}}}, \bibinfo {author}
  {\bibfnamefont {G.}~\bibnamefont {Birkl}}, \ and\ \bibinfo {author}
  {\bibfnamefont {M.}~\bibnamefont {Lewenstein}},\ }\href@noop {} {\bibfield
  {journal} {\bibinfo  {journal} {Phys.\,Rev.\,A}\ }\textbf {\bibinfo {volume}
  {66}},\ \bibinfo {pages} {042317} (\bibinfo {year} {2002})}\BibitemShut
  {NoStop}%
\bibitem [{\citenamefont {Strauch}\ \emph {et~al.}(2008)\citenamefont
  {Strauch}, \citenamefont {Edwards}, \citenamefont {Tiesinga}, \citenamefont
  {Williams},\ and\ \citenamefont {Clark}}]{cold:stra08}%
  \BibitemOpen
  \bibfield  {author} {\bibinfo {author} {\bibfnamefont {F.~W.}\ \bibnamefont
  {Strauch}}, \bibinfo {author} {\bibfnamefont {M.}~\bibnamefont {Edwards}},
  \bibinfo {author} {\bibfnamefont {E.}~\bibnamefont {Tiesinga}}, \bibinfo
  {author} {\bibfnamefont {C.}~\bibnamefont {Williams}}, \ and\ \bibinfo
  {author} {\bibfnamefont {C.~W.}\ \bibnamefont {Clark}},\ }\href@noop {}
  {\bibfield  {journal} {\bibinfo  {journal} {Phys.\,Rev.\,A}\ }\textbf
  {\bibinfo {volume} {77}},\ \bibinfo {pages} {050304} (\bibinfo {year}
  {2008})}\BibitemShut {NoStop}%
\bibitem [{\citenamefont {Eckardt}\ \emph {et~al.}(2005)\citenamefont
  {Eckardt}, \citenamefont {Weiss},\ and\ \citenamefont
  {Holthaus}}]{cold:ecka05}%
  \BibitemOpen
  \bibfield  {author} {\bibinfo {author} {\bibfnamefont {A.}~\bibnamefont
  {Eckardt}}, \bibinfo {author} {\bibfnamefont {C.}~\bibnamefont {Weiss}}, \
  and\ \bibinfo {author} {\bibfnamefont {M.}~\bibnamefont {Holthaus}},\
  }\href@noop {} {\bibfield  {journal} {\bibinfo  {journal}
  {Phys.\,Rev.\,Lett.}\ }\textbf {\bibinfo {volume} {95}},\ \bibinfo {pages}
  {260404} (\bibinfo {year} {2005})}\BibitemShut {NoStop}%
\bibitem [{\citenamefont {Lignier}\ \emph {et~al.}(2007)\citenamefont
  {Lignier}, \citenamefont {Sias}, \citenamefont {Ciampini}, \citenamefont
  {Singh}, \citenamefont {Zenesini}, \citenamefont {Morsch},\ and\
  \citenamefont {Arimondo}}]{cold:lign07}%
  \BibitemOpen
  \bibfield  {author} {\bibinfo {author} {\bibfnamefont {H.}~\bibnamefont
  {Lignier}}, \bibinfo {author} {\bibfnamefont {C.}~\bibnamefont {Sias}},
  \bibinfo {author} {\bibfnamefont {D.}~\bibnamefont {Ciampini}}, \bibinfo
  {author} {\bibfnamefont {Y.}~\bibnamefont {Singh}}, \bibinfo {author}
  {\bibfnamefont {A.}~\bibnamefont {Zenesini}}, \bibinfo {author}
  {\bibfnamefont {O.}~\bibnamefont {Morsch}}, \ and\ \bibinfo {author}
  {\bibfnamefont {E.}~\bibnamefont {Arimondo}},\ }\href@noop {} {\bibfield
  {journal} {\bibinfo  {journal} {Phys.\,Rev.\,Lett.}\ }\textbf {\bibinfo
  {volume} {99}},\ \bibinfo {pages} {220403} (\bibinfo {year}
  {2007})}\BibitemShut {NoStop}%
\bibitem [{\citenamefont {Vandersypen}\ and\ \citenamefont
  {Chuang}(2005)}]{cold:vand05}%
  \BibitemOpen
  \bibfield  {author} {\bibinfo {author} {\bibfnamefont {L.~M.~K.}\
  \bibnamefont {Vandersypen}}\ and\ \bibinfo {author} {\bibfnamefont {I.~L.}\
  \bibnamefont {Chuang}},\ }\href@noop {} {\bibfield  {journal} {\bibinfo
  {journal} {Rev.\,Mod.\,Phys.}\ }\textbf {\bibinfo {volume} {76}},\ \bibinfo
  {pages} {1037} (\bibinfo {year} {2005})}\BibitemShut {NoStop}%
\bibitem [{\citenamefont {Schneider}\ \emph {et~al.}(2009)\citenamefont
  {Schneider}, \citenamefont {Grishkevich},\ and\ \citenamefont
  {Saenz}}]{cold:schn09}%
  \BibitemOpen
  \bibfield  {author} {\bibinfo {author} {\bibfnamefont {P.-I.}\ \bibnamefont
  {Schneider}}, \bibinfo {author} {\bibfnamefont {S.}~\bibnamefont
  {Grishkevich}}, \ and\ \bibinfo {author} {\bibfnamefont {A.}~\bibnamefont
  {Saenz}},\ }\href@noop {} {\bibfield  {journal} {\bibinfo  {journal}
  {Phys.\,Rev.\,A}\ }\textbf {\bibinfo {volume} {80}},\ \bibinfo {pages}
  {013404} (\bibinfo {year} {2009})}\BibitemShut {NoStop}%
\bibitem [{Note1()}]{Note1}%
  \BibitemOpen
  \bibinfo {note} {Throughout the Letter we use a pulse envelope $f_{\protect
  \rm sh}(t)$ that for $0\leq t\leq t_{\protect \rm end}$ is defined by
  $H_2(\tau )e^{-\tau ^2}$ with $\tau = (1-2t/t_{\protect \rm end})/\protect
  \sqrt 2$ and $H_2$ the $2^{\protect \rm nd}$ Hermite polynomial.}\BibitemShut
  {Stop}%
\bibitem [{\citenamefont {M\"uller}\ \emph {et~al.}(2007)\citenamefont
  {M\"uller}, \citenamefont {F\"olling}, \citenamefont {Widera},\ and\
  \citenamefont {Bloch}}]{cold:muel07}%
  \BibitemOpen
  \bibfield  {author} {\bibinfo {author} {\bibfnamefont {T.}~\bibnamefont
  {M\"uller}}, \bibinfo {author} {\bibfnamefont {S.}~\bibnamefont {F\"olling}},
  \bibinfo {author} {\bibfnamefont {A.}~\bibnamefont {Widera}}, \ and\ \bibinfo
  {author} {\bibfnamefont {I.}~\bibnamefont {Bloch}},\ }\href@noop {}
  {\bibfield  {journal} {\bibinfo  {journal} {Phys.\,Rev.\,Lett.}\ }\textbf
  {\bibinfo {volume} {99}},\ \bibinfo {pages} {200405} (\bibinfo {year}
  {2007})}\BibitemShut {NoStop}%
\bibitem [{\citenamefont {Scully}\ and\ \citenamefont
  {Zubairy}(2002)}]{cold:scul02}%
  \BibitemOpen
  \bibfield  {author} {\bibinfo {author} {\bibfnamefont {M.~O.}\ \bibnamefont
  {Scully}}\ and\ \bibinfo {author} {\bibfnamefont {M.~S.}\ \bibnamefont
  {Zubairy}},\ }\href@noop {} {\emph {\bibinfo {title} {Quantum Optics}}}\
  (\bibinfo  {publisher} {Cambridge University Press},\ \bibinfo {address}
  {Cambridge},\ \bibinfo {year} {2002})\BibitemShut {NoStop}%
\bibitem [{\citenamefont {Bremner}\ \emph {et~al.}(2002)\citenamefont
  {Bremner}, \citenamefont {Dawson}, \citenamefont {Dodd}, \citenamefont
  {Gilchrist}, \citenamefont {Harrow}, \citenamefont {Mortimer}, \citenamefont
  {Nielsen},\ and\ \citenamefont {Osborne}}]{cold:brem02}%
  \BibitemOpen
  \bibfield  {author} {\bibinfo {author} {\bibfnamefont {M.~J.}\ \bibnamefont
  {Bremner}}, \bibinfo {author} {\bibfnamefont {C.~M.}\ \bibnamefont {Dawson}},
  \bibinfo {author} {\bibfnamefont {J.~L.}\ \bibnamefont {Dodd}}, \bibinfo
  {author} {\bibfnamefont {A.}~\bibnamefont {Gilchrist}}, \bibinfo {author}
  {\bibfnamefont {A.~W.}\ \bibnamefont {Harrow}}, \bibinfo {author}
  {\bibfnamefont {D.}~\bibnamefont {Mortimer}}, \bibinfo {author}
  {\bibfnamefont {M.~A.}\ \bibnamefont {Nielsen}}, \ and\ \bibinfo {author}
  {\bibfnamefont {T.~J.}\ \bibnamefont {Osborne}},\ }\href@noop {} {\bibfield
  {journal} {\bibinfo  {journal} {Phys.\,Rev.\,Lett.}\ }\textbf {\bibinfo
  {volume} {89}},\ \bibinfo {pages} {247902} (\bibinfo {year}
  {2002})}\BibitemShut {NoStop}%
\bibitem [{\citenamefont {Winkler}\ \emph {et~al.}(2006)\citenamefont
  {Winkler}, \citenamefont {Thalhammer}, \citenamefont {Lang}, \citenamefont
  {Grimm}, \citenamefont {Hecker-Denschlag}, \citenamefont {Daley},
  \citenamefont {Kantian}, \citenamefont {B\"uchler},\ and\ \citenamefont
  {Zoller}}]{cold:wink06}%
  \BibitemOpen
  \bibfield  {author} {\bibinfo {author} {\bibfnamefont {K.}~\bibnamefont
  {Winkler}}, \bibinfo {author} {\bibfnamefont {G.}~\bibnamefont {Thalhammer}},
  \bibinfo {author} {\bibfnamefont {F.}~\bibnamefont {Lang}}, \bibinfo {author}
  {\bibfnamefont {R.}~\bibnamefont {Grimm}}, \bibinfo {author} {\bibfnamefont
  {J.}~\bibnamefont {Hecker-Denschlag}}, \bibinfo {author} {\bibfnamefont
  {A.~J.}\ \bibnamefont {Daley}}, \bibinfo {author} {\bibfnamefont
  {A.}~\bibnamefont {Kantian}}, \bibinfo {author} {\bibfnamefont {H.~P.}\
  \bibnamefont {B\"uchler}}, \ and\ \bibinfo {author} {\bibfnamefont
  {P.}~\bibnamefont {Zoller}},\ }\href@noop {} {\bibfield  {journal} {\bibinfo
  {journal} {Nature}\ }\textbf {\bibinfo {volume} {441}},\ \bibinfo {pages}
  {853} (\bibinfo {year} {2006})}\BibitemShut {NoStop}%
\bibitem [{\citenamefont {Nielsen}(2002)}]{cold:niel02}%
  \BibitemOpen
  \bibfield  {author} {\bibinfo {author} {\bibfnamefont {M.~A.}\ \bibnamefont
  {Nielsen}},\ }\href@noop {} {\bibfield  {journal} {\bibinfo  {journal}
  {Phys.\,Lett.\,A}\ }\textbf {\bibinfo {volume} {303}},\ \bibinfo {pages}
  {249} (\bibinfo {year} {2002})}\BibitemShut {NoStop}%
\bibitem [{\citenamefont {Byrd}\ \emph {et~al.}(2005)\citenamefont {Byrd},
  \citenamefont {Lidar}, \citenamefont {Wu},\ and\ \citenamefont
  {Zanardi}}]{cold:byrd05}%
  \BibitemOpen
  \bibfield  {author} {\bibinfo {author} {\bibfnamefont {M.~S.}\ \bibnamefont
  {Byrd}}, \bibinfo {author} {\bibfnamefont {D.~A.}\ \bibnamefont {Lidar}},
  \bibinfo {author} {\bibfnamefont {L.-A.}\ \bibnamefont {Wu}}, \ and\ \bibinfo
  {author} {\bibfnamefont {P.}~\bibnamefont {Zanardi}},\ }\href@noop {}
  {\bibfield  {journal} {\bibinfo  {journal} {Phys.\,Rev.\,A}\ }\textbf
  {\bibinfo {volume} {71}},\ \bibinfo {pages} {052301} (\bibinfo {year}
  {2005})}\BibitemShut {NoStop}%
\bibitem [{\citenamefont {Niu}\ \emph {et~al.}(1996)\citenamefont {Niu},
  \citenamefont {Zhao}, \citenamefont {Georgakis},\ and\ \citenamefont
  {Raizen}}]{cold:niu96}%
  \BibitemOpen
  \bibfield  {author} {\bibinfo {author} {\bibfnamefont {Q.}~\bibnamefont
  {Niu}}, \bibinfo {author} {\bibfnamefont {X.-G.}\ \bibnamefont {Zhao}},
  \bibinfo {author} {\bibfnamefont {G.~A.}\ \bibnamefont {Georgakis}}, \ and\
  \bibinfo {author} {\bibfnamefont {M.~G.}\ \bibnamefont {Raizen}},\
  }\href@noop {} {\bibfield  {journal} {\bibinfo  {journal}
  {Phys.\,Rev.\,Lett.}\ }\textbf {\bibinfo {volume} {76}},\ \bibinfo {pages}
  {4504} (\bibinfo {year} {1996})}\BibitemShut {NoStop}%
\bibitem [{\citenamefont {Madison}\ \emph {et~al.}(1998)\citenamefont
  {Madison}, \citenamefont {Fischer}, \citenamefont {Diener}, \citenamefont
  {Niu},\ and\ \citenamefont {Raizen}}]{cold:madi98}%
  \BibitemOpen
  \bibfield  {author} {\bibinfo {author} {\bibfnamefont {K.~W.}\ \bibnamefont
  {Madison}}, \bibinfo {author} {\bibfnamefont {M.~C.}\ \bibnamefont
  {Fischer}}, \bibinfo {author} {\bibfnamefont {R.~B.}\ \bibnamefont {Diener}},
  \bibinfo {author} {\bibfnamefont {Q.}~\bibnamefont {Niu}}, \ and\ \bibinfo
  {author} {\bibfnamefont {M.~G.}\ \bibnamefont {Raizen}},\ }\href@noop {}
  {\bibfield  {journal} {\bibinfo  {journal} {Phys.\,Rev.\,Lett.}\ }\textbf
  {\bibinfo {volume} {81}},\ \bibinfo {pages} {5093} (\bibinfo {year}
  {1998})}\BibitemShut {NoStop}%
\bibitem [{\citenamefont {Bakr}\ \emph {et~al.}(2011)\citenamefont {Bakr},
  \citenamefont {Preiss}, \citenamefont {Tai}, \citenamefont {Ma},
  \citenamefont {Simon},\ and\ \citenamefont {Greiner}}]{cold:bakr11}%
  \BibitemOpen
  \bibfield  {author} {\bibinfo {author} {\bibfnamefont {W.~S.}\ \bibnamefont
  {Bakr}}, \bibinfo {author} {\bibfnamefont {P.~M.}\ \bibnamefont {Preiss}},
  \bibinfo {author} {\bibfnamefont {M.~E.}\ \bibnamefont {Tai}}, \bibinfo
  {author} {\bibfnamefont {R.}~\bibnamefont {Ma}}, \bibinfo {author}
  {\bibfnamefont {J.}~\bibnamefont {Simon}}, \ and\ \bibinfo {author}
  {\bibfnamefont {M.}~\bibnamefont {Greiner}},\ }\href@noop {} {\bibfield
  {journal} {\bibinfo  {journal} {Nature}\ }\textbf {\bibinfo {volume} {480}},\
  \bibinfo {pages} {500} (\bibinfo {year} {2011})}\BibitemShut {NoStop}%
\end{thebibliography}
%
\end{document}